\documentclass{prop2015}
\usepackage[english]{babel}
 \category{FQMT-2015}
\title{Quantum phase slips and voltage fluctuations in superconducting nanowires}
\author[A.\ G. Semenov]{Andrew G. Semenov\inst{1,3}}
\author[A.\ D. Zaikin]{Andrei D. Zaikin\inst{2,1,}\footnote{Corresponding author\quad E-mail:~\textsf{andrei.zaikin@kit.edu}}}
\address[1]{I.E.Tamm Department of Theoretical Physics, P.N.Lebedev
Physics Institute, 119991 Moscow, Russia}
\address[2]{Institute of Nanotechnology, Karlsruhe Institute of Technology (KIT), 76021 Karlsruhe, Germany}
\address[3]{National Research University Higher School of Economics, 101000 Moscow, Russia}
\begin{abstract}
We argue that quantum phase slips (QPS) may generate non-equilibrium voltage fluctuations in superconducting nanowires. In the low frequency limit we evaluate all cumulants of the voltage operator which obey Poisson statistics and show a power law dependence on the external bias. We specifically address quantum shot noise which power spectrum $S_\Omega$ may depend non-monotonously on temperature. In the long wire limit $S_\Omega$ decreases with increasing frequency $\Omega$ and vanishes beyond a threshold value of $\Omega$ at $T \to 0$. Our predictions can be directly tested in future experiments with superconducting nanowires.
\end{abstract}
\shortabstract
\begin{document}
\maketitle

\section{Introduction}

To conduct electric current without any resistance is the most fundamental property of any bulk superconducting material. Usually, the behavior of such superconductors is well described within the
framework of the standard mean field theory. The situation changes dramatically as soon as a superconducting structure (e.g., forming a narrow wire) becomes sufficiently thin. In such structures thermal and/or quantum fluctuations start playing an important role being responsible for temporal local suppression of the superconducting order parameter $\Delta=|\Delta|e^{i\varphi}$ inside the wire and, hence, for the phase slippage process. This process gives rise to novel physical phenomena which cannot adequately be described with the aid of the mean field theory.

In the low temperature limit thermal fluctuations are irrelevant and the system behavior is dominated by
quantum phase slips (QPS) \cite{AGZ,Bezr08,Z10,Bezrbook}.  Each QPS accounts for the net phase jump by
$\delta \varphi =\pm 2\pi$ implying a voltage pulse $\delta V=\dot{\varphi}/2e$
and tunneling of one magnetic flux quantum $\Phi_0\equiv \pi/e =\int |\delta V(t)|dt$ across the wire in the direction perpendicular to its axis. Formally such QPS events can be considered as quantum particles
interacting logarithmically between each other in space-time \cite{ZGOZ}. Accordingly, the ground state
of ultrathin superconducting wires can be described in terms of a 2d gas of interacting quantum phase slips with effective fugacity proportional to the QPS tunneling amplitude per unit wire length \cite{GZQPS}
\begin{equation}
\gamma_{QPS} \sim (g_\xi\Delta_0/\xi)\exp (-ag_\xi), \quad a \sim 1.
\label{gQPS}
\end{equation}
Here $g_\xi =2\pi\sigma_N s/(e^2\xi) \gg 1$ is the dimensionless normal state conductance of the
wire segment of length equal to the coherence length $\xi$, $\Delta_0$ is the mean field order parameter value, $\sigma_N$ and $s$ are respectively the wire Drude conductance and cross section.

At $T \to 0$ long superconducting wires suffer a quantum phase transition \cite{ZGOZ} governed by the dimensionless parameter $\lambda \propto \sqrt{s}$ (to be defined later). In ultrathin wires with $\lambda < 2$ superconductivity is completely destroyed by quantum fluctuations, and such wires can even go insulating
at $T=0$. In relatively thicker wires with $\lambda > 2$ quantum fluctuations are not so pronounced,
the wire resistance $R$ decreases with $T$ and one finds \cite{ZGOZ}
\begin{equation}
\label{Rdiff}
R=\frac{d\langle \hat V\rangle}{dI}
\propto
\begin{cases}
\gamma_{QPS}^2T^{2\lambda -3}, & T\gg \Phi_0I,
\\
\gamma_{QPS}^2I^{2\lambda -3}, & T\ll \Phi_0I.
\end{cases}
\end{equation}
Here and below $\langle \hat V\rangle$ denotes the expectation value of the voltage operator across the wire. According to this result the wire non-linear resistance does not vanish down to lowest temperatures, as it was indeed observed in many experiments  \cite{BT,Lau,Zgi08,liege}. The physics behind this result is transparent: An external current $I$ flowing along the wire breaks the symmetry between positive and negative voltage pulses making the former more likely than the latter. As a result, the net voltage drop $\langle \hat V\rangle$ occurs across the wire also implying non-zero resistance (\ref{Rdiff}).

Can one also expect to observe non-vanishing voltage fluctuations in superconducting nanowires?
The presence of QPS-induced {\it equilibrium} voltage fluctuations in
such nanowires can be predicted already on the basis of the result (\ref{Rdiff}) combined with the fluctuation-dissipation theorem. The issue of {\it non-equilibrium}
voltage fluctuations (e.g., shot noise) is more tricky. This issue requires a detailed theoretical analysis which is the main goal of the present paper.

Below we will demonstrate that non-equilibrium voltage fluctuations in ultrathin superconducting wires are caused by the process of quantum tunneling of magnetic flux quanta $\Phi_0$ which can be described by Poisson statistics. In particular, we will investigate
QPS-induced shot noise of the voltage in such wires and predict a highly non-trivial dependence of the noise power spectrum on temperature, frequency and external current.

\section{The system setup and the Hamiltonian}

\begin{figure}
  \includegraphics[width=\columnwidth]{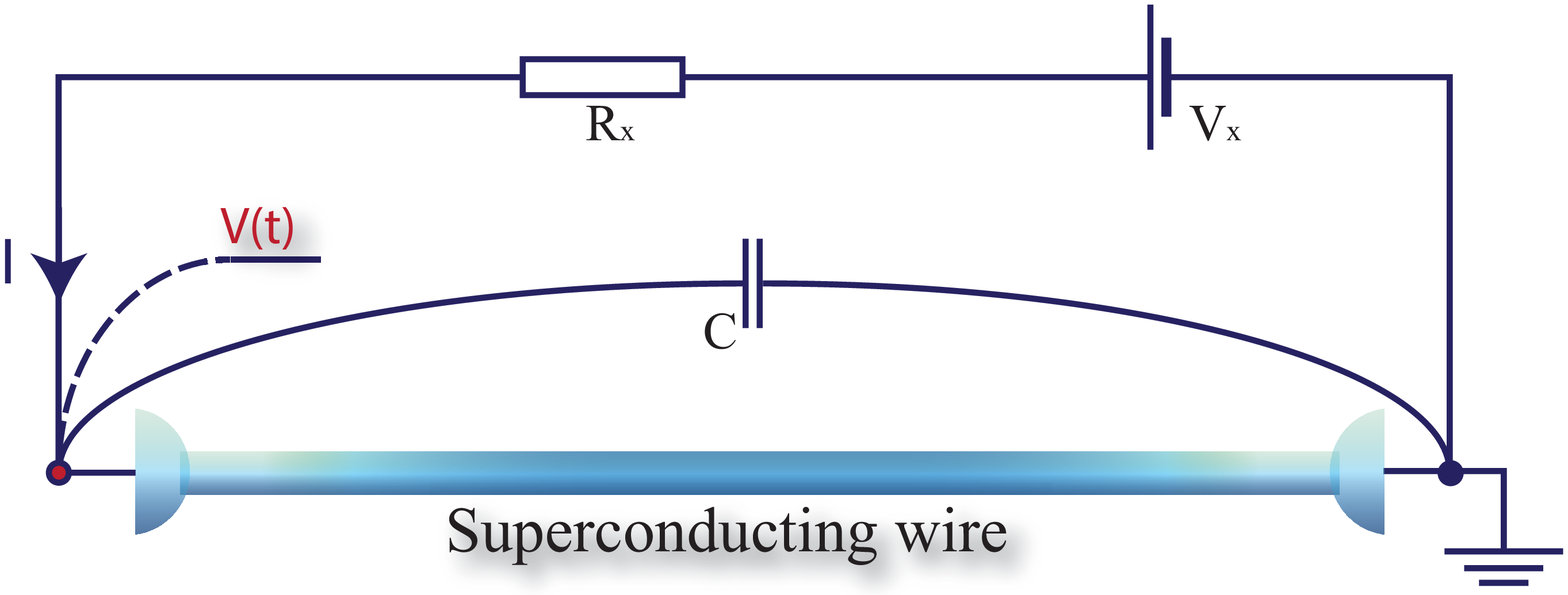}%
  \caption{\label{sysetup}\col
    The system under consideration.}
\end{figure}

In what follows we will consider an experimental setup depicted in Figure \ref{sysetup}. It consists of a thin superconducting wire of length $L$ and cross section $s$ attached to a voltage source $V_x$ by means of a  resistor $R_x$. A capacitor $C$ is switched in parallel to the superconducting wire. The right end of the wire is grounded as shown in the figure. Voltage fluctuations at its left end can be measured by a detector.

At low enough energies the superconducting wire can be described by an effective Lagrangian \cite{ZGOZ,GZQPS,ogzb}
\begin{equation}
\frac{1}{4e^2}\int dx\left(\frac{C_{\rm w}}{2}(\dot\varphi(x,t))^2-\frac{1}{2 \mathcal{L}_{\rm kin}}(\nabla\varphi(x,t))^2\right),
\label{Lvarphi}
\end{equation}
where $x$ is the coordinate along the wire ranging from 0 to $L$, $C_{\rm w}$ denotes geometric wire capacitance per unit length and $\mathcal{L}_{\rm kin}=1/(\pi\sigma_N\Delta_0 s)$ is the kinetic wire inductance (times length). Below we will employ the well known property enabling one to describe any electric circuit either in terms of the phases (node variables) or in terms of the transferred charges (loop variables), see, e.g., \cite{Zagoskin} for further details. It is remarkable that under certain conditions these two approaches can turn effectively dual to each other. The duality of this kind was established and discussed in details, e.g., for ultrasmall Josephson junctions \cite{PZ88,averin,Z90,SZ90}, as well in the case of short \cite{MN} and long \cite{SZ13} superconducting wires. According to the results \cite{SZ13} the dual representation for the Hamiltonian of a superconducting nanowire is defined by an effective sine-Gordon model
\begin{equation}
\hat H_{\rm wire} = \hat H_{TL}+ \hat H_{QPS}.
\label{Hamw}
\end{equation}
Here
\begin{equation}
\hat H_{ TL}=\int_0^Ldx \left(\frac{\hat \Phi^2(x)}{2\mathcal{L}_{\rm kin}}+\frac{(\nabla\hat \chi(x) )^2}{2C_{\rm w}\Phi_0^2}\right)
\end{equation}
defines the wire Hamiltonian in the absence of quantum phase slips. It describes an effective transmission line in terms of canonically conjugate flux (or phase) and charge operators obeying
the commutation relation
\begin{equation}
[\hat \Phi (x),\hat \chi (x')]=-i\Phi_0\delta (x-x').
\end{equation}
The term
\begin{equation}
\hat H_{QPS}=-\gamma_{QPS}\int_0^Ldx \cos(\hat \chi(x))
\label{HQPS}
\end{equation}
accounts for QPS effects in our nanowire.

Note that the quantum field $\chi (x,t)$ is proportional to the total charge $q(x,t)$ that has passed through
the point $x$ up to the time moment $t$, i.e. $q(x,t)=\chi (x,t)/\Phi_0$. Hence, the local current $I(x,t)$ and the local charge density $\rho (x,t)$ are defined by the equations
\begin{equation}
I(x,t)=\dot \chi (x,t)/\Phi_0, \quad \rho (x,t)=-\nabla\chi (x,t)/\Phi_0.
\end{equation}

In order to construct the total Hamiltonian $\hat H$ for our system it is also necessary to include the charging energy of a capacitor $C$ as well as the Hamiltonian of the external circuit. Below we will assume that the external resistor $R_x$ is very large, i.e. the wire is biased by a constant current $I=V_x/R_x$ which does not fluctuate. Then for the total Hamiltonian we have
\begin{equation}
\hat H = \hat H_{\rm wire}+\frac{(\hat Q-\hat \chi(0)/\Phi_0)^2}{2C}-\frac{I\hat\varphi}{2e}.
\end{equation}
Here the second and the third terms in the right-hand side represent respectively the charging energy and the potential energy tilt produced by the current $I$. The operator $\hat Q$ accounts for the charge across the capacitor and the phase operator $\hat\varphi$ corresponds to the variable $\varphi (t) \equiv \varphi (0,t)$ which represents the phase of the superconducting order parameter $\Delta (x,t)$ at $x=0$. Setting $\varphi (L,t)\equiv 0$, we conclude that the latter operator is related to the fluctuating voltage across the wire $V(t)$ as
\begin{equation}
\hat V(t)=\dot{\hat \varphi}/(2e).
\label{Jos}
\end{equation}

\section{Keldysh perturbation theory}
The task at hand is to investigate fluctuations of the voltage $V(t)$ in the presence of quantum phase slips.
We will proceed with the aid of the Keldysh path integral technique. As usually, we define our variables of interest on the forward and backward time parts of the Keldysh contour, $\varphi_{F,B} (t)$ and $\chi_{F,B}(x,t)$, and introduce the ``classical'' and ``quantum'' variables, respectively
$\varphi_+ (t)= (\varphi_F (t)+\varphi_B (t))/2$ and $\varphi_- (t)= \varphi_F (t)-\varphi_B (t)$ (and similarly for the $\chi$-fields).

Making use of Eq. (\ref{Jos}), let us express the general correlator of voltages in the form
\begin{multline}
\langle V(t_1)V(t_2)...V(t_n)\rangle =\frac{1}{(2e)^n}   \\
\times\left\langle\dot \varphi_{+}(t_1)\dot \varphi_{+}(t_2)... \dot\varphi_{+}(t_n) e^{iS_{QPS}}\right\rangle_0,
\label{VVn}
\end{multline}
where
\begin{equation}
 S_{QPS}=-2 \gamma_{QPS}\int dt\int\limits_0^L dx\sin(\chi_{+})\sin (\chi_-/2)
\end{equation}
and
\begin{equation}
\langle ...\rangle_0 =\int\mathcal D^2\varphi (t) \mathcal D^2\chi (x,t) (...) e^{iS_0[\varphi , \chi ]}
\end{equation}
indicates averaging with the effective action $S_0[\varphi , \chi ]$  corresponding to the Hamiltonian $\hat H_{0} = \hat H- \hat H_{QPS}$.

It is important to emphasize that Eq. (\ref{VVn}) defines the symmetrized voltage correlators. E.g., for $n=2$ one has
\begin{equation}
\langle V(t_1)V(t_2)\rangle =\frac12\langle\hat V(t_1)\hat V(t_2)+\hat V(t_2)\hat V(t_1)\rangle,
\end{equation}
while for $n=3$ one can verify that \cite{3rdcum}

\begin{multline}
\langle V(t_1)V(t_2)V(t_3)\rangle =\frac{1}{8}
 \big\{\langle\hat V(t_1)\big({\cal T}\hat V(t_2)\hat V(t_3)\big)\rangle\\
+\langle\big(\tilde{\cal T}\hat V(t_2)\hat V(t_3)\big)\hat V(t_1)\rangle
+\,\langle\hat V(t_2)\big({\cal T}\hat V(t_1)\hat V(t_3)\big)\rangle\\
+\langle\big(\tilde{\cal T}\hat V(t_1)\hat V(t_3)\big)\hat V(t_2)\rangle
+\,\langle\hat V(t_3)\big({\cal T}\hat V(t_1)\hat V(t_2)\big)\rangle\\
+\langle\big(\tilde{\cal T}\hat V(t_1)\hat V(t_2)\big)\hat V(t_3)\rangle
+\,\langle{\cal T}\hat V(t_1)\hat V(t_2)\hat V(t_3)\rangle\\
+\langle\tilde{\cal T}\hat V(t_1)\hat V(t_2)\hat V(t_3)\rangle\big\},
\label{tri}
\end{multline}
where ${\cal T}$ and $\tilde{\cal T}$ are, respectively, the forward and backward time ordering operators.

Equation (\ref{VVn}) is a formally exact expression which we will now evaluate perturbatively in the tunneling amplitude $\gamma_{QPS}$ (\ref{gQPS}). In the zero order in $\gamma_{QPS}$ the problem is described by the quadratic (in both $\varphi$ and $\chi$) action $S_0$. In that case it is necessary to employ the averages
\begin{gather}
\langle\varphi_+(t)\rangle_0=\langle\varphi_-(t)\rangle_0=\langle\chi_-(x,t)\rangle_0=0, \nonumber\\
\langle\chi_+(x,t)\rangle_0=\Phi_0It,
\end{gather}
as well as pair averages (the Green functions):
\begin{gather}
G^K_{ab}(X,X')=-i\langle a_{+}(X)b_{+}(X')\rangle_0 +i \langle a_{+}(X)\rangle_0\langle b_{+}(X')\rangle_0 , \nonumber\\
G^R_{ab}(X,X')=-i\langle a_{+}(X)b_-(X')\rangle_0 ,
\label{GRK}
\end{gather}
where $a(X)$ and $b(X)$ stand for one of the fields $\varphi (t)$ and $\chi (x,t)$. As both these fields are
real, the advanced  and retarded Green functions satisfy the condition $G^A_{ab}(\omega)=G^R_{ba}(-\omega)$. Due to linearity the Keldysh function $G^K$ can then be expressed in the form
\begin{eqnarray}
 G^K_{ab}(\omega)=\frac12\coth\left(\frac{\omega}{2T}\right)\left(G^R_{ab}(\omega)-G^R_{ba}(-\omega)\right)
\label{GK}\end{eqnarray}
even in presence of the external bias $V_x$. For the same reason the retarded Green function can be evaluated either from the full quantum mechanical treatment or just from simple electrotechnical arguments.
For the function $G^R_{\varphi\varphi}$ we obtain
\begin{equation}
G^R_{\varphi\varphi}(\omega)=\frac{1}{\frac{\omega^2}{2E_C}+\frac{ i\omega}{4e^2R_x}-\frac{\omega\lambda}{\pi}\cot\left(\frac{\omega L}{v}\right)},
\label{Gvv}
\end{equation}
where $E_C=e^2/(2C)$, $v=1/\sqrt{{\mathcal L}_{\rm kin}C_{\rm w}}$ is the plasmon velocity \cite{ms}
and the parameter $\lambda$ already introduced above is defined as $\lambda=R_Q/(2Z_{\rm w})$ with $R_Q =\pi/(2e^2)$ being the "superconducting" quantum resistance unit and $Z_{\rm w}=\sqrt{\mathcal{L}_{\rm kin}/C_{\rm w}}$ being the wire impedance.

The corresponding expressions for $G^R_{\varphi\chi}$ and $G^R_{\chi\chi}$ turn out somewhat lengthy. In order to simplify them it is useful to bear in mind that due to momentum conservation plasmons in our system can only be created in pairs with the total zero momentum. All plasmons moving towards the grounded end of the wire eventually disappear and never pop up again while excitations moving in the opposite direction produce voltage fluctuations measured by a detector. Then in the interesting for us long wire limit the general expressions for $G^R_{\varphi\chi}$ and $G^R_{\chi\chi}$ reduce to
\begin{eqnarray}
G^R_{\varphi\chi}(x;\omega)\simeq -\frac{2\lambda e^{i\frac{\omega x}{v}}}{(\omega+ i0)\left(\frac{\omega}{2E_C}+\frac{i\lambda}{\pi}\right)}, \label{GvcGc}\\
G^R_{\chi\chi}(x,x';\omega)\simeq -\frac{2\pi i\lambda }{\omega+i0}e^{i\frac{\omega|x-x'|}{v}}.
\label{GvcGcc}
\end{eqnarray}
In Eqs. (\ref{GvcGc}) and (\ref{GvcGcc}) we also set $R_x\to\infty$ as requested in the current bias limit.

\begin{figure}
\includegraphics[width=\columnwidth]{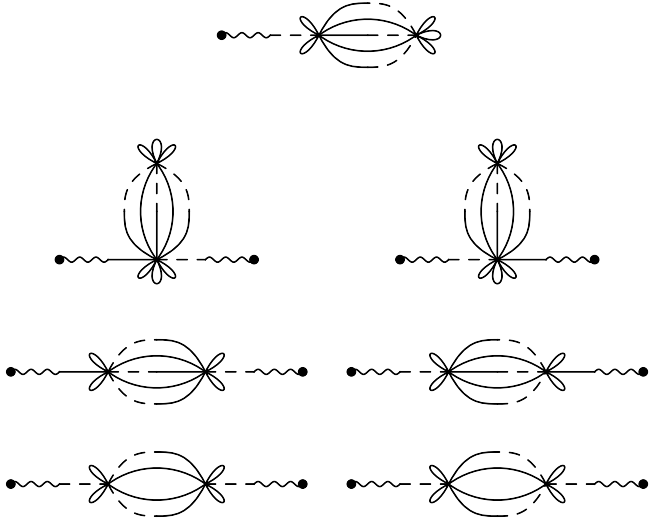}
\caption{Candy-like diagrams which determine both average voltage $\langle V \rangle$ (upper diagram) and voltage-voltage corrrelator $\langle VV \rangle$ (six remaining diagrams) in the second order in $\gamma_{QPS}$. The fields $\varphi_+$, $\chi_+$ and $\chi_-$ in the propagators (\ref{GRK}) are denoted respectively by wavy, solid and dashed lines.}
\label{candys}
\end{figure}

Expanding Eq. (\ref{VVn}) up to the second order in $\gamma_{QPS}$ and performing all necessary averages,
we evaluate the results in terms of the Green functions (\ref{GRK}). Pictorially these results can also be represented in the form of the so-called candy diagrams \cite{SZ16}. These diagrams for the first and the second moments of the voltage operator are displayed in Figure \ref{candys}.

\section{Average voltage}

To begin with, let us evaluate the expectation value of the voltage operator. Our perturbation yields
\begin{multline}
 \langle \hat V\rangle=\frac{i\gamma_{QPS}^2}{8e}\int\limits_{0}^L dx\int\limits_0^L dx'\left(\lim_{\omega\to 0}\omega G_{\varphi\chi}^R(x;\omega)\right)
 \\
 \times\left({\mathcal P}_{x,x'}(-\Phi_0 I)-{\mathcal P}_{x,x'}(\Phi_0I)\right),
\label{V1}
\end{multline}
where ${\mathcal P}_{x,x'}(\omega)=P_{x,x'}(\omega)+\bar P_{x,x'}(\omega)$ and
\begin{equation}
 \label{P}
 P_{x,x'}(\omega)=\int\limits_{0}^\infty dt e^{i\omega t}e^{i{\mathcal G}(x,x';t,0)-\frac{i}{2}{\mathcal G}(x,x;0,0)-\frac{i}{2}{\mathcal G}(x',x';0,0)},
\end{equation}
 \begin{equation}
{\mathcal G}(x,x';t,0)= G^K_{\chi\chi}(x,x';t,0)+\frac 12 G^R_{\chi\chi}(x,x';t,0).
\end{equation}
Making use of the identity $\lim_{\omega\to 0}\omega G_{\varphi\chi}^R(x;\omega)=2\pi i$, we can reduce Eq. (\ref{V1}) to a simple form
\begin{equation}
\langle \hat V \rangle =\Phi_0\left(\Gamma_{QPS}(I)-\Gamma_{QPS}(-I)\right),
\label{V2}
\end{equation}
where
\begin{equation}
 \Gamma_{QPS}(I)=\frac{\gamma_{QPS}^2}{4}\int\limits_{0}^L dx\int\limits_0^L dx'{\mathcal P}_{x,x'}(\Phi_0I).
\label{GQPS}
\end{equation}
Employing the above results and making use of the generalized detailed balance condition
\begin{equation}
{\mathcal P}_{x,x'}(\omega)=e^{\frac{\omega}{T}}{\mathcal P}_{x,x'}(-\omega),
\label{dbc}
\end{equation}
we obtain
\begin{equation}
\langle V\rangle =\frac{ vL \gamma_{QPS}^2\Phi_0}{4}\varsigma^2\left(\frac{\Phi_0 I}{2}\right)\sinh\left(\frac{\Phi_0 I}{2T}\right),
\label{V3}
\end{equation}
where
\begin{equation}
\varsigma(\omega)=\tau_0^\lambda (2\pi T)^{\lambda -1}\frac{\Gamma\left(\frac{\lambda}{2}-\frac{i\omega}{2\pi T}\right)\Gamma\left(\frac{\lambda}{2}+\frac{i\omega}{2\pi T}\right)}{\Gamma(\lambda)},
\label{vsi}
\end{equation}
$\tau_0 \sim 1/\Delta_0$ is the QPS core size in time and $\Gamma (x)$ is the
Euler Gamma-function. The result (\ref{V3}), (\ref{vsi}) yields Eq. (\ref{Rdiff}) in the corresponding limits and matches with the analogous expression derived in \cite{ZGOZ} by means of a different technique.

\section{Relation to Im$F$-method}
Comparing Eq. (\ref{V2}) with the corresponding result for the average voltage \cite{ZGOZ} we immediately conclude that the quantity $\Gamma_{QPS}(I)$ (\ref{GQPS}) can be interpreted as a quantum decay rate of the current state due to QPS. In \cite{ZGOZ} this rate was evaluated from the imaginary
part of the free energy by means of the so-called ${\rm Im}F$-method, see, e.g., \cite{Weiss}. It is instructive to establish a detailed relation between the latter approach and the Keldysh technique employed here.

To this end let us define the generalized Green function $\mathcal G_{\chi}(x,x';\sigma)$ which depends on the complex time $\sigma$ and obeys the condition $\mathcal G_\chi(x,x';t-i0)=\mathcal G(x,x';t,0)$ at $t>0$.
It reads
\begin{multline}
\mathcal G_{\chi}(x,x';\sigma) = \frac{iT}{2}\int dt \coth(\pi T(t-\sigma))\\
\times\int\frac{d\omega}{2\pi} e^{-i\omega t}\left(G^{R}_{\chi\chi}(x,x';\omega)-G^{R}_{\chi\chi}(x,x';-\omega)\right)
\end{multline}
This function is analytic, has branch cuts at ${\rm Im}(\sigma)=N/T$ for all integer $N$ and is periodic in the imaginary time, i.e.
\begin{equation}
\mathcal G_{\chi}(x,x';\sigma)=\mathcal G_{\chi}(x,x';\sigma-i/T).
\label{periodic}
\end{equation}
Note that Eq. (\ref{periodic}) just follows from the Kubo-Martin-Schwinger condition. On the imaginary axis  the function $\mathcal G_{\chi}$ coincides with the Matsubara Green function
\begin{equation}
\mathcal G_{\chi}(x,x';-i\tau)=i G_{\chi\chi}^M(x,x';\tau).
\label{mats}
\end{equation}
\begin{figure}
\includegraphics[width=\columnwidth]{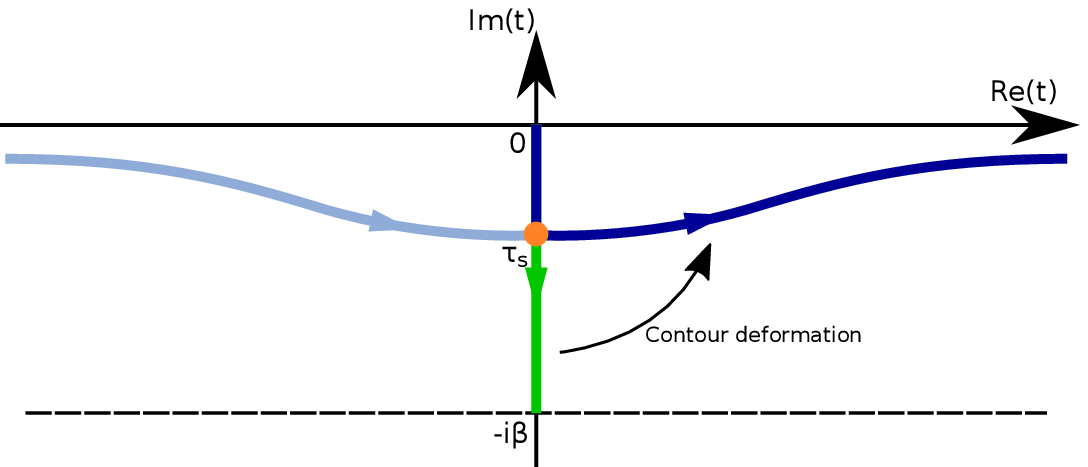}
\caption{\col Integration contour.}
\label{analytic}
\end{figure}
Evaluating the quantum decay rate $\Gamma$ by means of the pioneered by Langer ${\rm Im}F$-method one employs a general formula
\begin{equation}
   \Gamma =-2{\rm Im}F,
\label{ImF}
\end{equation}
where $F$ is the system free energy. In order to establish the QPS contribution to $\Gamma$ it is necessary
to evaluate the corresponding correction to the free energy $\delta F$. In the leading order in $\gamma_{QPS}$ it suffices to consider just one QPS-anti-QPS pair \cite{ZGOZ} which yields
\begin{equation}
\delta F\approx-\frac{\gamma_{QPS}^2}{4}\int\limits_{0}^L dx\int\limits_0^L dx' \int\limits_0^{1/T} d\tau e^{-S_{pair}},
\label{deltaF}
\end{equation}
where
\begin{equation}
S_{pair}=-\Phi_0I\tau+\mathcal V(x,\tau;x',0),
\end{equation}
where $\tau$ is the imaginary time interval between QPS and anti-QPS events and $\mathcal V(x;x';\tau,0)$ describes the interaction between these two events occuring respectively at the points $x$ and $x'$. This interaction term can be expressed via the Matsubara Green function as
\begin{multline}
\mathcal V(x;x';\tau,0) = G_{\chi\chi}^M(x,x';\tau)\\-\frac12G_{\chi\chi}^M(x,x;0)-\frac12G_{\chi\chi}^M(x',x';0).
\label{iaiint}
\end{multline}
Note that the integral over $\tau$ in Eq. (\ref{deltaF}) is formally divergent at low temperatures. As a result, the free energy acquires an imaginary part Im$F$ derived by means of a proper analytic continuation of $\delta F$. Evaluating the integral (\ref{deltaF}) by the steepest descent method we first determine a stationary point $\tau_s$ from the stationary condition for the action
\begin{equation}
\Phi_0I=\partial_\tau G_{\chi\chi}^M(x,x';\tau_s).
\end{equation}
A closer inspection demonstrates that this stationary point delivers a maximum to the action rather than a minimum, thereby indicating an instability with respect to QPS-mediated decay to lower energy states. In this case the correct recipe is to deform the integration contour along the steepest descent path. This procedure
is illustrated in Figure \ref{analytic}. The initial integration path is a vertical line going from $0$ to $-i\beta$ (note that the Matsubara technique operates with imaginary times).  The deformed contour is directed along the real-time axis after passing through the point $\tau_s$. With this in mind we obtain
\begin{multline}
\delta F\approx-\frac{\gamma_{QPS}^2}{4}\int\limits_{0}^L dx\int\limits_0^L dx' \int\limits_0^{\tau_s} d\tau e^{\Phi_0 I\tau -\mathcal V(x;x';\tau,0)}\\
-\frac{\gamma_{QPS}^2}{4}\int\limits_{0}^L dx\int\limits_0^L dx' \int\limits_{0}^\infty id\tau e^{\Phi_0 I(\tau_s+i\tau) -\mathcal V(x;x';\tau_s+i\tau,0)}.
\end{multline}
Taking the imaginary part of this expression, we get
\begin{multline}
-2{\rm Im} F= \frac{\gamma_{QPS}^2}{4}\int\limits_{0}^L dx\int\limits_0^L dx' \int\limits_{0}^\infty d\tau e^{\Phi_0 I(\tau_s+i\tau) -\mathcal V(x;x';\tau_s+i\tau,0)}\\+\frac{\gamma_{QPS}^2}{4}\int\limits_{0}^L dx\int\limits_0^L dx' \int\limits_{0}^\infty d\tau e^{\Phi_0 I(\tau_s-i\tau) -\mathcal V(x;x';\tau_s-i\tau,0)}.
\label{IFIF}
\end{multline}
Further expressing  Eq. (\ref{IFIF}) as a single integral along the contour passing through the point $\tau_s$ in the direction perpendicular to real $\tau$ axis, we find
\begin{equation}
-2{\rm Im} F=\frac{\gamma_{QPS}^2}{4}\int\limits_{0}^L dx\int\limits_0^L dx' \int\limits_{-\infty}^\infty d\tau e^{\Phi_0 I(\tau_s+i\tau) -\mathcal V(x;x';\tau_s+i\tau,0)}.
\end{equation}
Combined with Eqs. (\ref{iaiint}) and (\ref{mats}), this expression can also be rewritten in the form
\begin{multline}
-2{\rm Im} F=\frac{\gamma_{QPS}^2}{4}\int\limits_{0}^L dx\int\limits_0^L dx' \int\limits_{-\infty}^\infty dt e^{\Phi_0 I(\tau_s+it)}\\
\times e^{i{\mathcal G}_{\chi}(x,x';t-i\tau_s,0)-\frac{i}{2}{\mathcal G}_{\chi}(x,x;0,0)-\frac{i}{2}{\mathcal G}_{\chi}(x',x';0,0)}.
\end{multline}
Employing the relation
$$
\mathcal P_{x,x'}(\omega)=\int\limits_{-\infty}^\infty dt e^{i\omega t}e^{i\mathcal G_{\chi}(x,x';t-i0)-\frac{i}{2}{\mathcal G}_{\chi}(x,x;0,0)-\frac{i}{2}{\mathcal G}_{\chi}(x',x';0,0)},
$$
we arrive at the final result
\begin{equation}
-2{\rm Im} F=\frac{\gamma_{QPS}^2}{4}\int\limits_{0}^L dx\int\limits_0^L dx'{\mathcal P}_{x,x'}(\Phi_0I).
\end{equation}
This result together with Eq. (\ref{ImF}) demonstrates that Eq. (\ref{V2}) indeed defines the decay rate of the current states in a superconducting nanowire due to QPS, thereby proving the equivalence of the Im$F$-approach employed in \cite{ZGOZ} and the real time Keldysh technique combined with duality arguments elaborated here. The latter technique appears much more convenient for the analysis of voltage fluctuations
to be developed below.

\section{Voltage fluctuations}
Let us investigate the second moment of the voltage operator, i.e. the voltage noise.
Our perturbative analysis allows to recover three different contributions
to the noise power spectrum, i.e.
\begin{equation}
 S_{\Omega}=\int dt e^{i\Omega t}\langle V(t)V(0)\rangle =S^{(0)}_{\Omega}+S^{r}_{\Omega}+S^{a}_{\Omega}.
\label{VV2}
\end{equation}
The first of these contributions
\begin{equation}
S^{(0)}_{\Omega}= \frac{i\Omega^2\coth\left(\frac{\Omega}{2T}\right)}{16e^2}\left(G_{\varphi\varphi}^R(\Omega)-
G_{\varphi\varphi}^R(-\Omega)\right)
 \end{equation}
defines equilibrium voltage noise for a transmission line and has nothing to do with QPS. The remaining two contributions are due to QPS effects. The term $S^{r}_{\Omega}$ contains the products of two retarded (advanced) Green functions:
\begin{multline}
S^{r}_{\Omega}=\frac{\gamma_{QPS}^2\Omega^2\coth\left(\frac{\Omega}{2T}\right)}{16e^2}\int\limits_{0}^L dx\int\limits_0^L dx' {\rm Re}\left[G_{\varphi\chi}^R(x;\Omega) \right.\\
 \left.\times ({\mathcal F}_{x,x'}(\Omega )G_{\varphi\chi}^R(x';\Omega)
 -{\mathcal F}_{x,x'}(0)G_{\varphi\chi}^R(x;\Omega))\right],
 \label{Sr}
 \end{multline}
where
\begin{multline}
{\mathcal F}_{x,x'}(\Omega )=
- P_{x,x'}(\Omega +\Phi_0I)-P_{x,x'}(\Omega -\Phi_0I)\\
+\bar P_{x,x'}(-\Omega +\Phi_0I)+\bar P_{x,x'}(- \Omega -\Phi_0I).
\end{multline}
The term $S^{a}_{\Omega}$, in contrast, contains the product of retarded and advanced Green functions
and reads
 \begin{eqnarray}
\label{Sa}
S^{a}_{\Omega}=\frac{\gamma_{QPS}^2\Omega^2}{32e^2}\int\limits_{0}^L dx\int\limits_0^L dx'G_{\varphi\chi}^R(x;\Omega)G_{\varphi\chi}^R(x';-\Omega)\\
\times \left[\sum_{\pm}{\mathcal Q}_{\pm}\left(
 {\mathcal P}_{x,x'}(\Omega\pm\Phi_0I)-{\mathcal P}_{x,x'}(-\Omega\mp\Phi_0I)\right)\right].\nonumber
\end{eqnarray}
Here we denoted
\begin{equation}
{\mathcal Q}_{\pm} = \coth\left(\frac{\Omega\pm\Phi_0 I}{2T}\right)-\coth\left(\frac{\Omega}{2T}\right).
\label{QQQ}
\end{equation}
Eqs. (\ref{VV2})-(\ref{QQQ}) together with the expressions for the Green functions (\ref{Gvv})-(\ref{GvcGcc}) fully determine the voltage noise power spectrum of a superconducting nanowire in the perturbative in QPS regime.

At non-zero bias values the QPS noise turns non-equilibrium. In the zero frequency limit $\Omega \to 0$
the terms $S^{(0)}_{\Omega}$ and $S^{r}_{\Omega}$ tend to zero, and the voltage noise $S_{\Omega \to 0}\equiv S_0$ is determined solely by $S^{a}_{\Omega}$. Then from Eq. (\ref{Sa}) we obtain
\begin{equation}
 S_{0}= \Phi_0^2\left(\Gamma_{QPS}(I)+\Gamma_{QPS}(-I)\right)
 =\Phi_0\coth\left(\frac{\Phi_0 I}{2T}\right)\langle V \rangle,
\label{shot1}
\end{equation}
where $\langle V \rangle$ is defined in Eqs. (\ref{V2}), (\ref{V3}).  In the low temperature limit $T\ll \Phi_0I$ Eq. (\ref{shot1}) accounts for QPS-induced {\it shot noise} $S_0=\Phi_0\langle V \rangle$ obeying {\it Poisson statistics} with an effective ``charge'' equal to the flux quantum $\Phi_0$.
In other words, shot noise in superconducting nanowires is produced by quantum tunneling of the magnetic flux across the wire. In the dual picture tunneling flux quanta $\Phi_0$ can be viewed as charged quantum particles passing through (and being scattered at) an effective "tunnel barrier" which role is played by the nanowire.

Let us also note that shot noise can also be generated in the high temperature regime dominated by thermally activated phase slips (TAPS) \cite{GZTAPS}. In the latter case this shot noise is again described by Eq. (\ref{shot1}) with the TAPS decay rate substituted instead of the QPS one, i.e. $\Gamma_{QPS}(I) \to \Gamma_{TAPS}(I)$.

Our analysis also allows to recover higher correlators of the voltage operator (\ref{VVn}). Let us define the
voltage cumulants
\begin{equation}
\mathcal C_n = (-i)^n\lim_{t\to\infty}\left[\frac{1}{t}\partial_z^n\log\left\langle e^{iz\int\limits_0^tdt_1V(t_1)} \right\rangle\right]_{z=0}
\label{Cn}
\end{equation}
In fact, within the accuracy of our analysis the terms ${\mathcal C}_k \propto \gamma_{QPS}^2$ with $k<n$ generated in the right-hand side of Eq. (\ref{Cn}) can be safely dropped and ${\mathcal C}_n$ just coincides with the Fourier transformed correlators (\ref{VVn}), i.e. ${\mathcal C}_2=S_0$ etc. As before, proceeding perturbatively in $\gamma_{QPS}$ and making use of Eqs. (\ref{V3}), (\ref{vsi}), at $T\to 0$ we find
\begin{equation}
{\mathcal C}_n=\Phi_0^{n-1}\langle V \rangle=\frac{\pi^2 vL\gamma^2_{QPS}\tau_0^{2\lambda}\Phi_0^{n}}{2^{2\lambda -2}\Gamma^2(\lambda)}|\Phi_0I|^{2\lambda -2}.
\end{equation}

Let us now consider voltage fluctuations at non-zero frequencies. Below we will restrict our analysis to voltage noise and stick to the limit of sufficiently high frequencies and/or long wires $v/L\ll\Omega\ll \Delta_0$. In this case the term $S^{(0)}_{\Omega}$ turns out to be independent of the wire length $L$. Evaluating the QPS terms $S^{r}_{\Omega}$ and $S^{a}_{\Omega}$, we observe that the latter scales linearly with the wire length $L$ whereas the former does not. Hence,
the terms $S^{(0)}_{\Omega}$ and $S^{r}_{\Omega}$ can be safely neglected in the long wire limit. For the remaining QPS contribution $S^{a}_{\Omega}$ we get
\begin{multline}
S^{a}_{\Omega}=\frac{vL\lambda^2\gamma_{QPS}^2}{8e^2}\left[\varsigma\left(\frac{\Phi_0 I}{2}-\Omega\right)
-\varsigma\left(\frac{\Phi_0 I}{2}+\Omega\right)\right]\\
\times\frac{\sinh\left(\frac{\Phi_0 I}{2T}\right)\varsigma\left(\frac{\Phi_0 I}{2}\right)}{\left((\Omega/2E_C)^2+(\lambda /\pi)^2\right)\sinh\left(\frac{\Omega}{2T}\right)}.
\label{Sa3}
\end{multline}
At $T\to 0$ from Eq. (\ref{Sa3}) we find
\begin{equation}
S^{a}_{\Omega}\propto
\begin{cases}
I^{\lambda -1}(I-2\Omega /\Phi_0)^{\lambda -1}, & \Omega < \Phi_0I/2,
\\
0, & \Omega > \Phi_0I/2.
\end{cases}
\label{Sa4}
\end{equation}
This result is explained as follows. At $T=0$ each QPS event can in general excite 2$N$ plasmons ($N=1,2...$) with total energy $E=\Phi_0 I$ and total zero momentum. $N$ plasmons (carrying total energy $E/2$) propagate towards the grounded end of the wire and eventually get dissipated there,
while the remaining $N$ plasmons (also with total energy $E/2$) propagate in the opposite direction reaching the opposite wire end and causing voltage fluctuations (emit a photon) with frequency $\Omega$ measured by a detector. At $T=0$ this process is only possible at $\Omega < E/2$ in the agreement with Eq. (\ref{Sa4}).

\begin{figure}
  \includegraphics[width=\columnwidth]{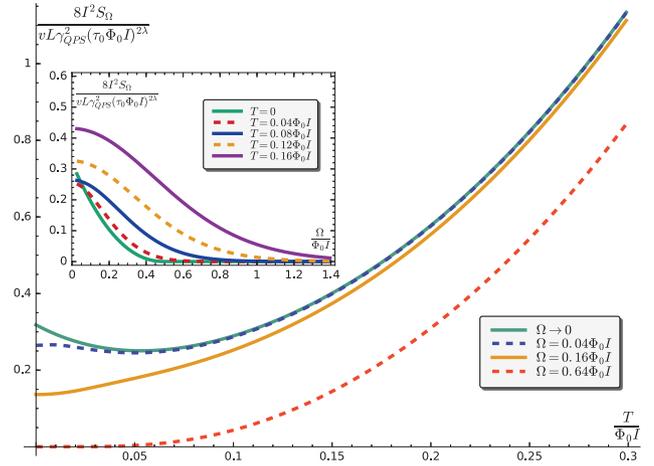}%
  \caption{\label{result}\col
    The temperature dependence of the QPS noise spectrum $S_{\Omega}$ (\ref{Sa3}) at $ \lambda =3.2$, large $E_C$ and different $\Omega$ in the long wire limit. The inset shows $S_{\Omega}$ as a function of $\Omega$ at different temperatures.}
\end{figure}

The result (\ref{Sa3}) is also illustrated in Figure 4. At sufficiently small $\Omega$ (we still keep $\Omega \gg v/L$) one observes a non-monotonous dependence of $S_{\Omega}$ on $T$ which is a direct consequence of quantum coherent nature of QPS noise. Note, that at non-zero $T$ the expression (\ref{Sa3})
{\it does not} coincide with the zero frequency result (\ref{shot1}) even in the limit $\Omega \to 0$.
The point here is that before taking the zero frequency limit in Eq. (\ref{Sa3}) one should formally set
$L \to \infty$. Then one gets
\begin{equation}
S_{\Omega\to 0}^{a}(I)=-\frac{vLT\gamma_{QPS}^2\Phi_0^2}{2}\varsigma\left(\frac{\Phi_0I}{2}\right)\varsigma'\left(\frac{\Phi_0I}{2}\right)\sinh\left(\frac{\Phi_0 I}{2T}\right).
 \label{s0o0}
\end{equation}
Comparing this expression with Eq. (\ref{shot1}) (combined with Eqs. (\ref{V3}), (\ref{vsi})) obtained in the true zero frequency limit (meaning that the limit $\Omega \to 0$ was taken prior to sending the wire length $L$ to infinity) one can establish the identity
\begin{equation}
S_0(I,T)-S_{\Omega\to 0 }^{a}(I,T)=2TR(I,T).
\end{equation}
It follows immediately that both expressions (\ref{s0o0}) and
(\ref{shot1}) coincide only at $T=0$, while at any non-zero $T$ the noise power $S_0(I,T)$ (\ref{shot1}) exceeds one in Eq. (\ref{s0o0})
and -- in contrast to the latter -- increases monotonously with temperature.

Finally, we point out that the perturbative in $\gamma_{QPS}$ approach employed here is applicable for
not too thin wires with $\lambda > 2$, i.e. in the "superconducting" phase. In thinner wires with $\lambda < 2$ characterized by unbound QPS-anti-QPS pairs (a non-superconducting phase) the QPS amplitude $\gamma_{QPS}$ gets effectively renormalized to higher values \cite{ZGOZ} and, hence,
the perturbation theory becomes obsolete in the low energy limit. However, even in this case our results may still remain applicable at sufficiently high temperature, frequency and/or current values.

This work was supported by the Russian Science Foundation under grant No. 16-12-10521.

\end{document}